\documentclass[showpacs,amssymb,floatfix]{revtex4}
\usepackage{graphicx}
\usepackage{dcolumn}
\usepackage{bm}
\begin{document}
\title{Two-electron bound states in continuum in quantum dots}
\author{A.F. Sadreev$^1$ and T.V. Babushkina$^2$}
\address{$^1$Institute of Physics, Academy of Sciences, 660036 Krasnoyarsk,
Russia\\ $^2$Siberian Federal University, Krasnoyarsk 660079
Krasnoyarsk}
\date{\today}
\begin{abstract}

Bound state in continuum (BIC) might appear in open quantum dots
for variation of the dot's shape. By means of the equations of
motion of Green functions we investigate effect of strong intradot
Coulomb interactions on that phenomenon in the framework of
impurity Anderson model. Equation that imaginary part of poles of
the  Green function equals to zero gives condition for BICs. As a
result we show that Coulomb interactions replicate the
single-electron BICs into two-electron ones.
\end{abstract}
\maketitle

In 1929, von Neumann and Wigner \cite{neumann} predicted the
existence of discrete solutions of the single-particle
Schr\"odinger equation embedded in the continuum of positive
energy states. Their analysis examined by Stillinger and Herrick
\cite{stillinger} long time was regarded as mathematical curiosity
because of certain spatially oscillating central symmetric
potentials.  More recently in 1973 Herrik \cite{herrik} and
Stillinger \cite{stillinger2} predicted BICs in semiconductor
heterostructure superlattices which were observed by Capasso {\it
et al} as the very narrow absorption peak \cite{capasso}.

 In the framework of the Feshbach's theory of
resonances Friedrich and Wintgen \cite{friedrich} have shown that
BIC occurs due to the interference of resonances. If two
resonances pass each other as a function of a continuous
parameter, then for a given value of the parameter one resonance
has exactly vanishing width. Later this result was reproduced in
application to different physical systems in the two-level
approximation \cite{shahbazyan,fanPRL,volya,guevara,RS1,PRB}.
Straight waveguide with an attractive, finite size impurity
presents an example of realistic structure in which Kim {\it et
al} presented the numerical evidence for the BIC for the variation
of the impurity size \cite{kim}. Further, calculations in
microwave and semiconductor open structures showed that the
resonance width also can turn to zero for variation of angle of
bent waveguide \cite{olendski}, shape of quantum dot (or
resonator) \cite{PRB,na}, or magnetic field \cite{ring}. Recently
it was rigorously shown that the zero resonance width is the
necessary and sufficient condition for BIC \cite{ring,PRA}. This
condition means that a coupling of the resonance state with
continuum equals zero to convert the state into BIC
\cite{texier,ring}.

That very restricted list of references shows that BICs might
occur in different open quantum systems including, for example,
laser induced continuum structures in atoms \cite{magunov1}, in
the molecular system \cite{cederbaum}. However in application to
open quantum dots (QD) BICs were studied in the single electron
approximation whereas the Coulomb interactions between electrons
might be very important for robustness of the BIC. In the present
work we consider effect of local intradot Coulomb interactions in
QD onto BICs in the framework of the two-level impurity Anderson
model \cite{anderson} that is one of the most important
theoretical models for a study of strong correlations in condensed
matter physics.

We consider quantum dot (QD) coupled to leads (left and right)
which support one propagating mode (the case of two continuums)
with the following total Hamiltonian
\begin{equation}\label{Htotal}
  H=\sum_{C=L,R}H_C+H_D+V.
\end{equation}
The leads, left (L) and right (R) in (\ref{Htotal}) are presented
as the non interacting electron gas
\begin{equation}\label{wires}
  H_C=\sum_{k\sigma}\epsilon(k)c_{k\sigma C}^{+}c_{k\sigma
  C}, ~C=L,R.
\end{equation}
A continual spectrum $\epsilon(k)$ defines the
propagating band of leads. The Hamiltonian of many level QD is
that of the impurity Anderson model \cite{anderson},
\begin{equation}\label{QD}
  H_D=\sum_{m\sigma}\epsilon_m a_{m\sigma}^{+}a_{m\sigma}+
\sum_{mn}U_{m}n_{m\sigma}n_{m\overline{\sigma}}.
\end{equation}
Here $a_{m\sigma}^{+}$ is the creation operator of an electron on
the m-th level of QD, $U_{m}$ takes into account the Hubbard
repulsion at the level $m$, and
$n_{m\sigma}=a_{m\sigma}^{+}a_{m\sigma}$. The interaction
\begin{equation}\label{V}
  V=\sum_{k\sigma m C}V_m(k)(c_{k\sigma C}^{+}a_{m\sigma}+h.c)
\end{equation}
describes couplings between the leads and QD where $c_{k\sigma C
}^{+}$ is the creation operator of an electron in the leads $C$.

In order to calculate transport properties of the QD we use a
technique of the equations of motion for retarded and advanced
Green functions which successfully used to consider the Fano and
Kondo resonances in the Anderson model
\cite{meir,jauho,lu,rudzinski,meden,trocha}. Following Laxroix
\cite{lacroix} we use a Hartree-Fock approximation in the wires
$\langle\langle c_{k\sigma
C}a_{n\overline{\sigma}}^{+}a_{n\overline{\sigma}}|a_{m\sigma'}^{+}
\rangle\rangle\approx \langle
n_{n\overline{\sigma}}\rangle\langle\langle c_{k\sigma
C}|a_{m\sigma'}^{+} \rangle\rangle $. The approximation is
justified for weak couplings compared to the Coulomb interactions:
$V_m\ll U_m$. As a result we obtain the following equation
\begin{equation}\label{Dyson}
  {\bf G}^{-1}(E)={\bf G}_{QD}^{-1}(E)+i{\bf \Gamma}
\end{equation}
for the Green functions $G_{m\sigma,n\sigma'}(E)=\langle\langle
a_{m\sigma}|a_{n\sigma'}^{+} \rangle\rangle^{-1}$ in the form of
the Dyson equation \cite{lu}. Here ${\bf G}_{QD}(E)$ is the Green
function of the isolated QD
\begin{eqnarray}\label{QDGF}
  &G_{QD,mm',\sigma,\sigma'}(E)=G_{QD,m\sigma}(E)\delta_{mm'}
  \delta_{\sigma,\sigma'}&\nonumber\\
  &G_{QD,m\sigma}(E)=\frac{1-\langle
  n_{m\overline{\sigma}}\rangle}{E-\epsilon_m}+
  \frac{\langle
  n_{m\overline{\sigma}}\rangle}{E-\epsilon_m-U_m}.&
\end{eqnarray}
which are exact for the isolated QD. For the simplicity  we take
wide band wires and approximate the self-energy as
\cite{hewson,lu}
\begin{equation}\label{Sigma1}
\sum_k\frac{V_m(k)V_n(k)}{E-\epsilon(k)_{\sigma}+i0}= -i\pi
V_mV_n\rho_C(E)=-i\sqrt{\Gamma_m\Gamma_n}
\end{equation}
where $\rho_C(E)$ is the density of states of the left and right
wires. The average values of the occupation numbers $\langle
n_{m\sigma}\rangle=\langle a_{m\sigma}^{+}a_{m\sigma}\rangle$
which enter the expressions for the Green functions are calculated
self-consistently via the formulas \cite{lacroix}
\begin{equation}\label{occup}
\langle n_{m\sigma}\rangle=\frac{1}{\pi}\int dE Im
G_{m\sigma,m\sigma}(E).
\end{equation}

The form of the self-energy (\ref{Sigma1}) and the QD Green
function (\ref{QDGF}) allows to proceed to the case of free
electrons with $U_m=0$. In this case BIC appears if QD acquires
accidental degeneracy $\epsilon_1=\epsilon_2$ \cite{PRB}.
Therefore in the vicinity of the degeneracy point
$\varepsilon=\epsilon_2-\epsilon_1=0$ we can restrict ourselves to
the two-level approximation for QD \cite{volya}. Then the
occupation numbers (\ref{occup}) are given by four poles of the
Green function (\ref{Dyson}).  At zero temperature, the
transmission amplitude can be expressed in term of the Green
function
\begin{equation}\label{T}
  T=\Gamma G(E)\Gamma^{+}, ~\Gamma=(\Gamma_1, \Gamma_2).
\end{equation}

\begin{figure}[ht]
\includegraphics[width=0.36\columnwidth]{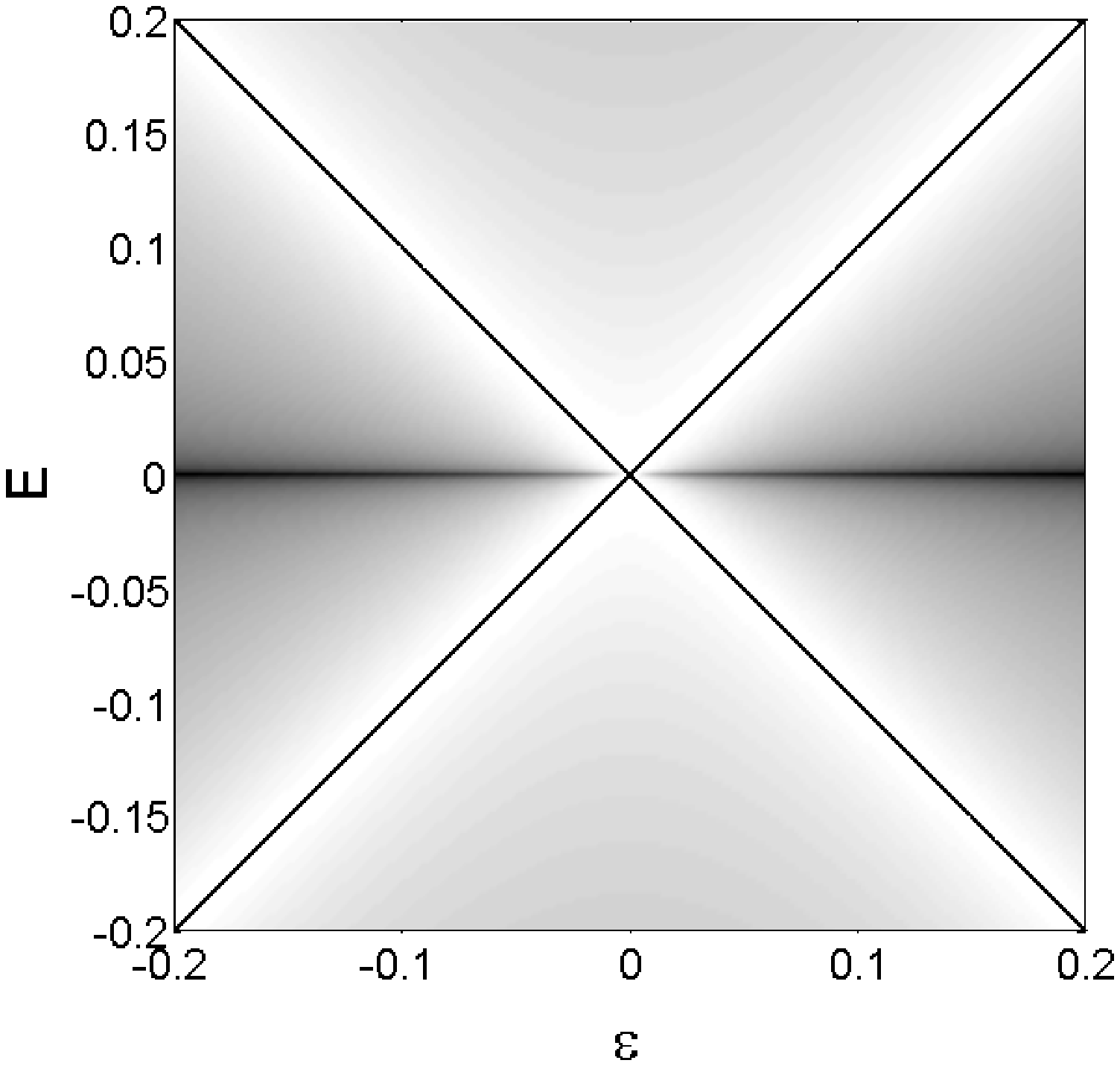}
\includegraphics[width=0.4\columnwidth]{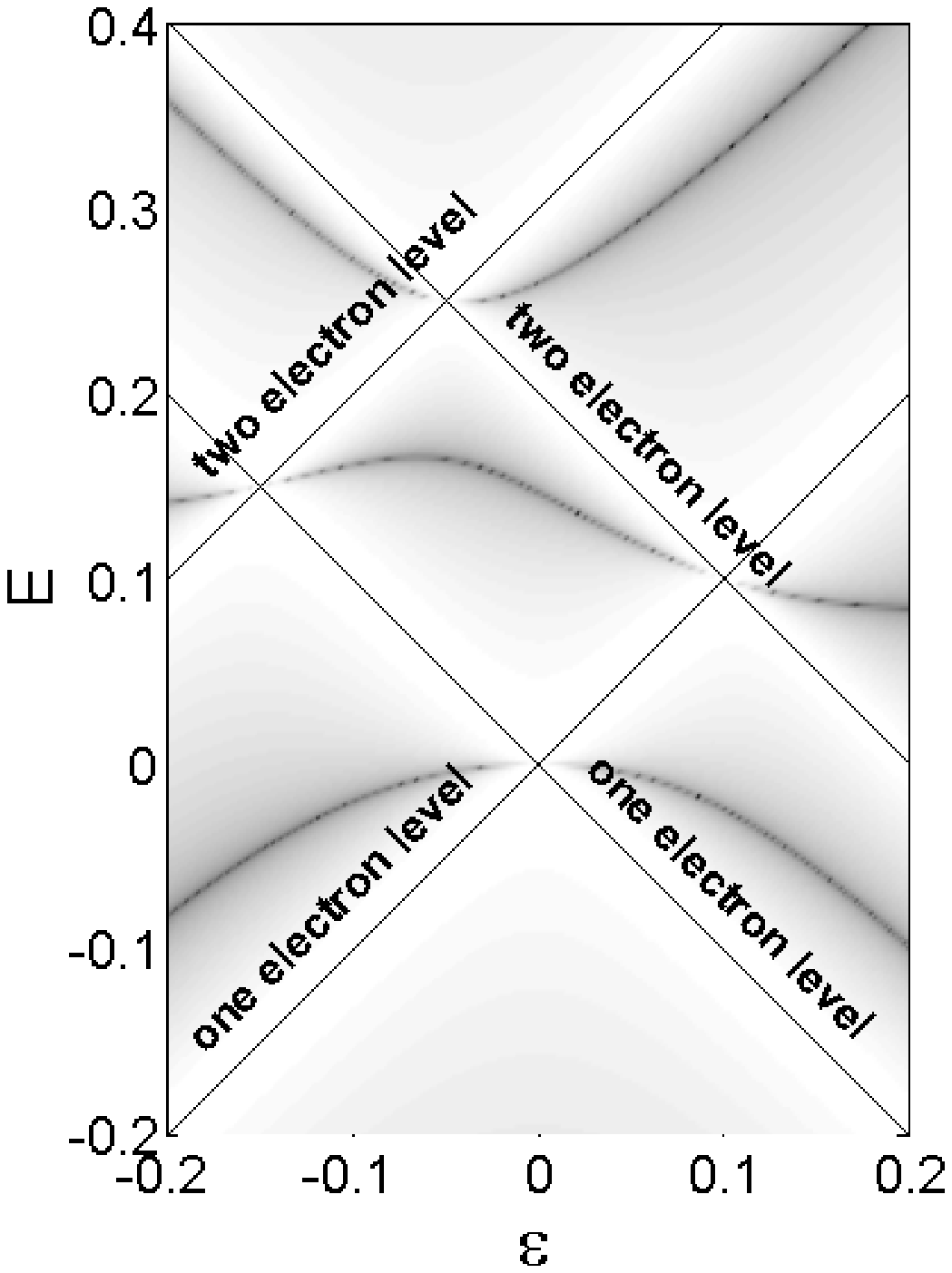}
\caption{(a) The transmission $\ln|T| $ of QD versus energy of
incident electron and energy splitting $\epsilon$ for the case of
zero Hubbard repulsion $U_1=U_2=0$. (b) The case of strong Hubbard
repulsion $U_1=0.2, ~U_2=0.3, \Gamma_1=\Gamma_2=0.05$. The
single-electron and two-electron energy levels in closed QD are
shown by thin lines. Black regions correspond to those where the
transmission close to zero while the white regions do to the
maximal transmission.} \label{fig1}
\end{figure}

\begin{figure}[ht]
\includegraphics[width=.4\columnwidth]{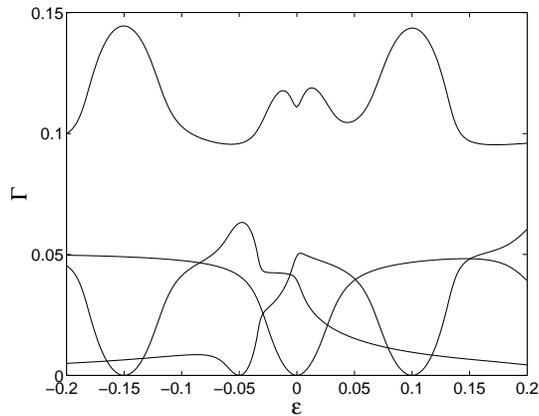}
\caption{The resonance widths defined as
$-2Im[z_{\lambda}(E,\epsilon)], ~\lambda=1, 2, 3, 4$ versus
$\epsilon$ for $E=0$ where $z_{\lambda}$ are the poles of the
Green function (\ref{Dyson}).} \label{fig2}
\end{figure}

The results of numerical self-consistent calculation of the
transmission (\ref{T}) are presented in Fig. \ref{fig1}. Fig.
\ref{fig1} (a) shows the case of zero Hubbard repulsion $U_{mn}=0$
(no electron correlations) in which QD is given only by the
single-electron energy levels. As shown in \cite{volya,PRB} BIC
occurs here at the point of degeneracy of electron states in QD
for $\epsilon=0$. At this point the S-matrix becomes singular
because the transmission zero crosses the unit transmission
\cite{PRB} as shown in Fig. \ref{fig1} where the unit transmission
follows the energy levels. As the Hubbard repulsion is included,
QD is given not only by single-electron states but also by
two-electron states as shown in Fig. \ref{fig1} (b) by solid
lines. As a result we obtain that the number of degenerated points
becomes four as seen from Fig. \ref{fig1} (b). One can see that
lines of zero transmission cross the lines of maximal unit
transmission at these points. Therefore, one can expect the BICs
at four points of degeneracy of the QD. In order to show that we
present in Fig. \ref{fig2} the resonance widths of the energy
levels defined as $\Gamma_{\lambda}=-2Im(z_{\lambda}), ~\lambda=1,
2, 3, 4$, where $z_{\lambda}(E,\epsilon)$ are the poles of the
Green function or zeros of the right hand expression in the Dyson
equation (\ref{Dyson}). The points at which $\Gamma_{\lambda}=0$
define BICs \cite{ring,PRA}. One can see that these points
coincide with the points of degeneracy of the QD given by
equations $\epsilon_{c1}=0,~
\epsilon_{c2}=U_1/2,~\epsilon_{c3}=-U_2/2,
~\epsilon_{c4}=(U_2-U_1)/2$. The corresponding energies of BICs
are $0, ~0.1, ~-0.15, 0.05$. The first BIC is pure single electron
localized state superposed of two single electron states with
$m=1,2$ \cite{PRB}. However the next two BICs are supespositions
of the single electron states and two-electron ones. At last, for
the last case ($\epsilon_{c4}=(U_2-U_1)/2$0 the BIC is superposed
of the two-electron states in QD. Although specific values of
$\Gamma_m$ has no importance for BIC's points defined by crossings
of the energy levels of QD, they are important for appearance of
the Dicke superradiant state which accumulates the total width
\cite{volya} as seen from Fig. \ref{fig2}.

\begin{figure}[ht]
\includegraphics[width=.3\columnwidth]{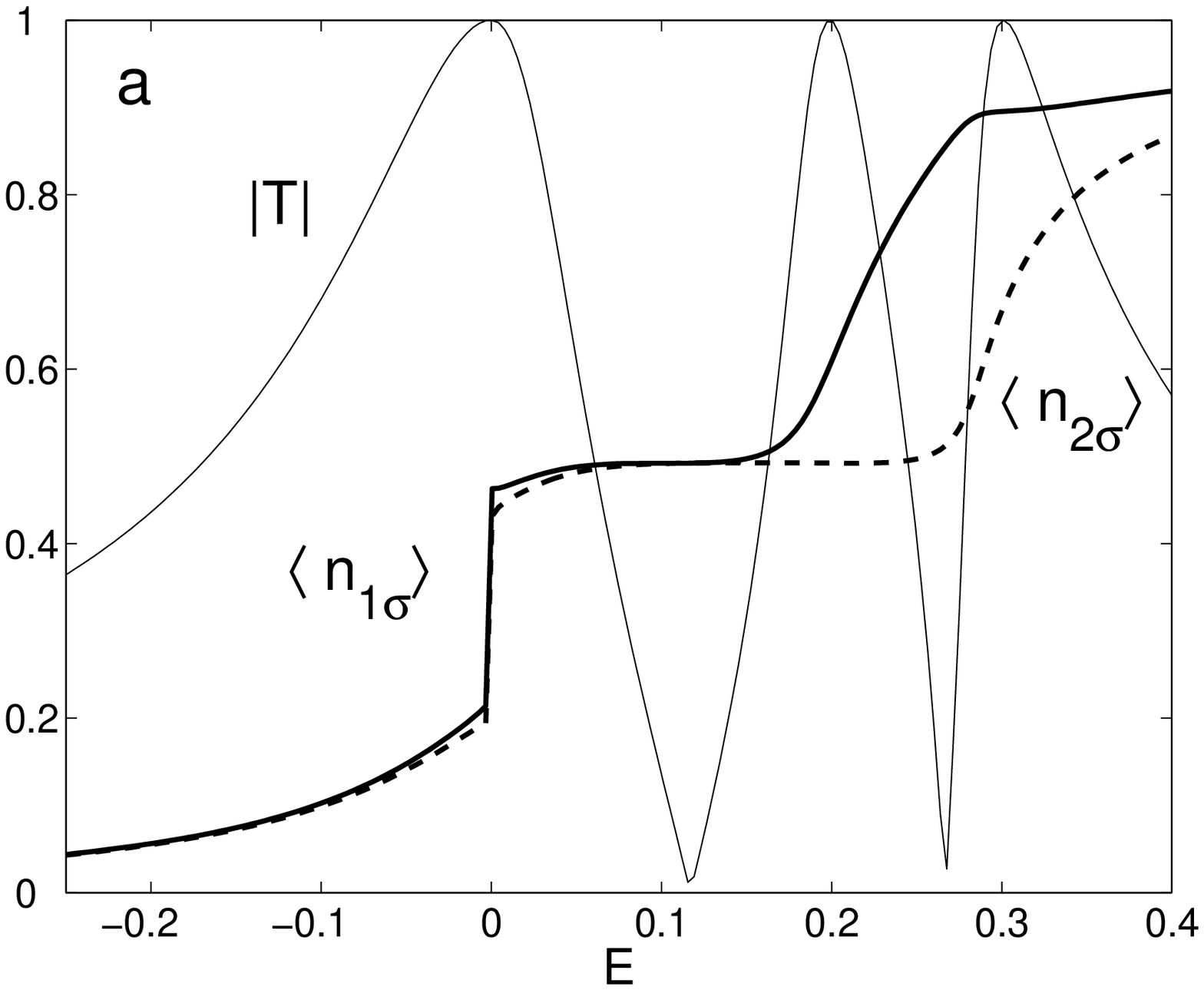}
\includegraphics[width=.3\columnwidth]{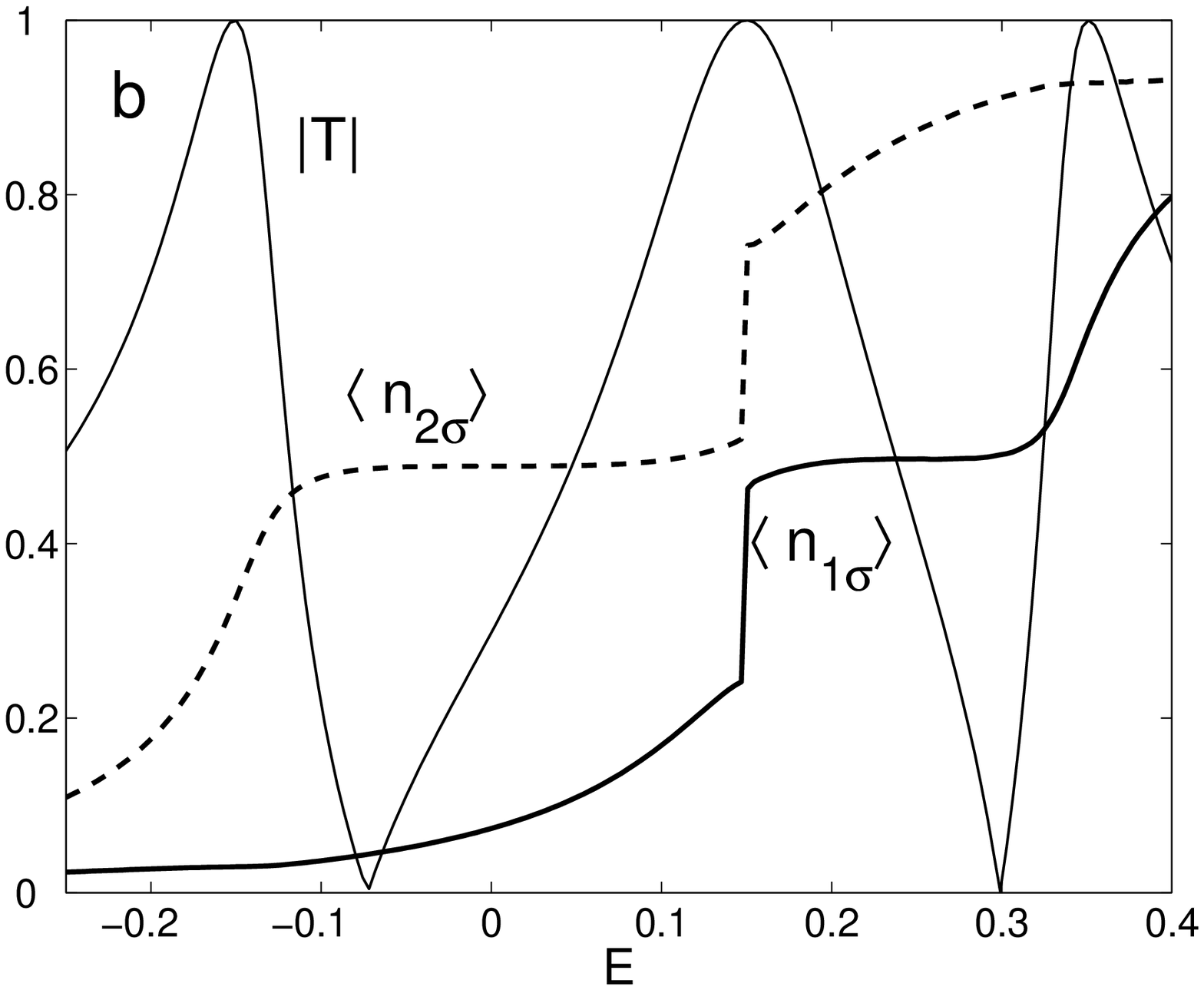}
\includegraphics[width=.31\columnwidth]{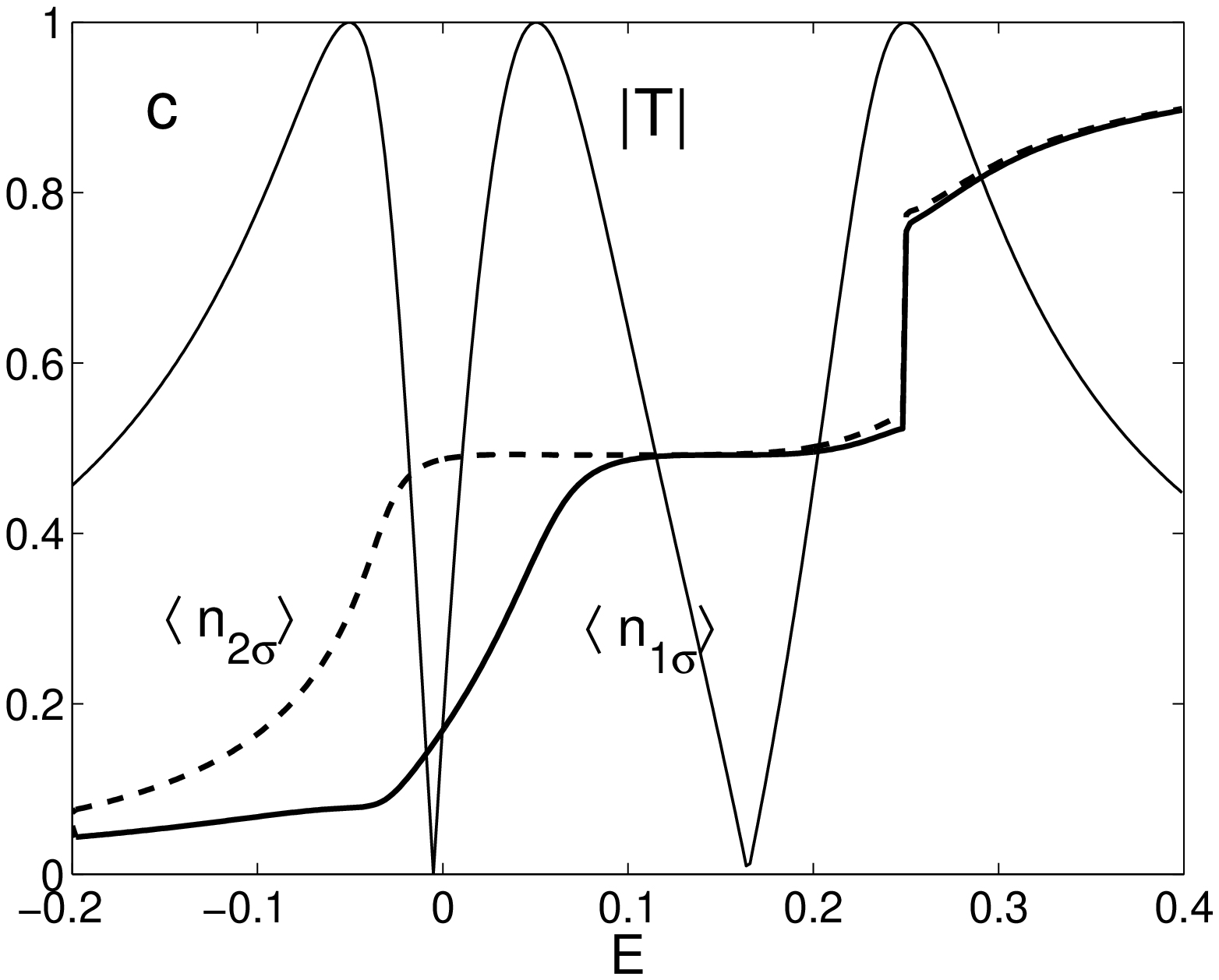}
\caption{The electron populations as dependent on the energy of
incident electron defined by (\ref{occup}) for the parameters of
the system given in Fig. \ref{fig1}. (a) $\epsilon=0$, (b)
$\epsilon=-0.15$, and (c) $\epsilon=-0.05$. The transmission $|T|$
is shown by thin green line.} \label{fig3}
\end{figure}

Since the resonance width turns to zero with approaching the BIC
point, we expect singular behavior of occupation numbers
(\ref{occup}) at the energy of BIC. In fact, Fig. \ref{fig2} (a,
b, c) demonstrate this effect. Let us consider the first BIC at
$\epsilon=0$ with discrete energy $E=0$ at which the
single-electron energies in QD are crossing as shown in Fig.
\ref{fig1} (b). One can see from Fig. \ref{fig3} (a) that at the
energy E=0 both energy levels are sharply and simultaneously
populated till one half. The next resonances with finite widths
correspond to the two-electron energies of QD  that are populated
smoothly at the Hubbard repulsive energies $U_1=0.2$ and $U_2=0.3$
by usual scenario as seen from Fig. \ref{fig3} (a).

The next BICs happen for the single-electron state crosses the
two-electron state at points $\epsilon=-0.15$ and $\epsilon=0.1$.
Because of similarity of these BIC points we have presented here
only the first case as shown in Fig. \ref{fig3} (b). The BIC's
discrete energy for that case equals to $E=0.15$ (Fig.
\ref{fig1}). Again we see that for approaching to this energy the
BIC populates sharply. However the populations of the
single-electron level and two-electron one are well separated
because of the Hubbard repulsion of the two-electron state. The
last figure Fig. \ref{fig3} (c) refers to the crossing of
two-electron energies at the point $\epsilon=-0.05$. As seen from
Fig. \ref{fig1} (a) the two-electron BIC has energy $E=0.25$. As a
result for approaching this energy we observe sharp population of
this state similar to the case (a).

Are BICs critical to energy level crossing? Similar to
\begin{figure}[ht]
\includegraphics[width=.4\columnwidth]{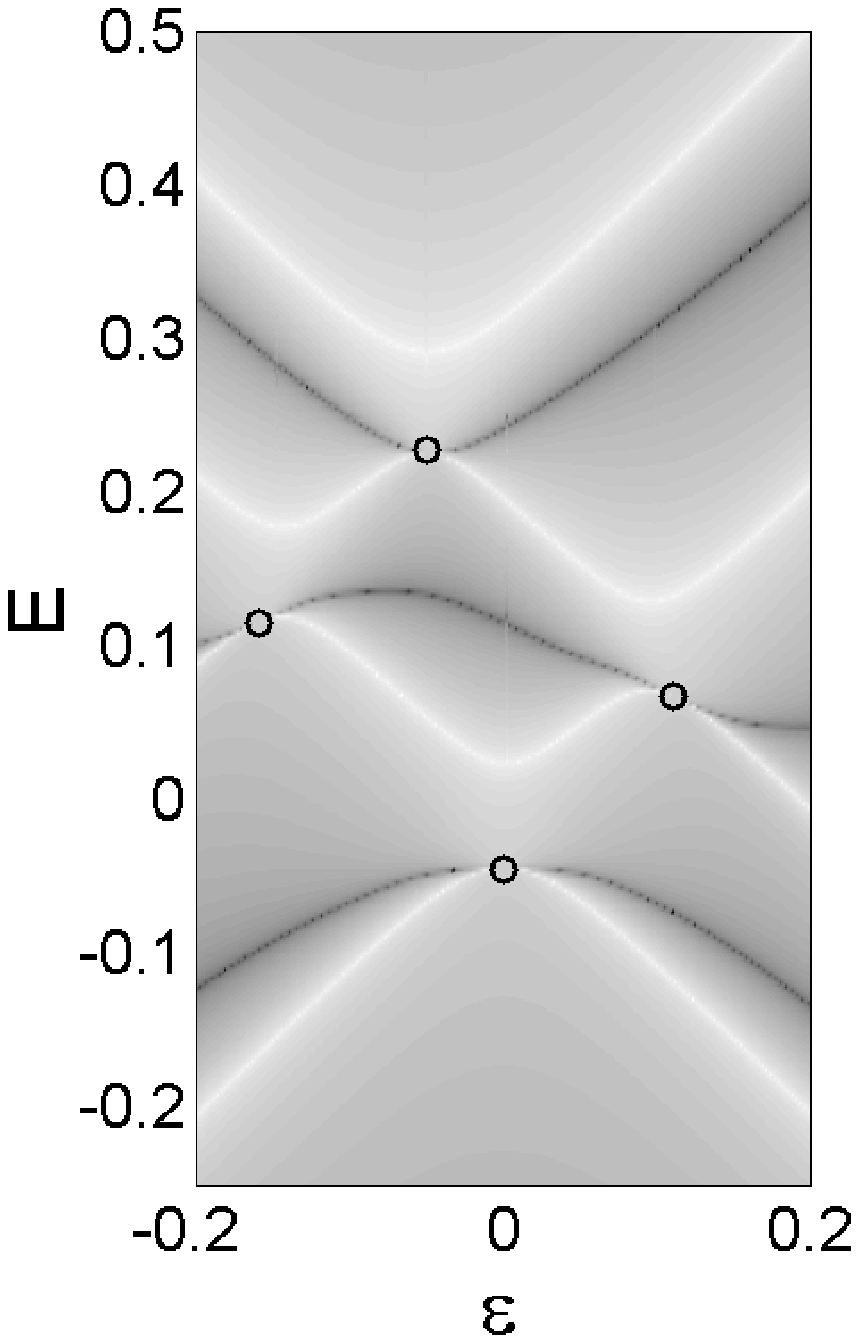}
\includegraphics[width=.38\columnwidth]{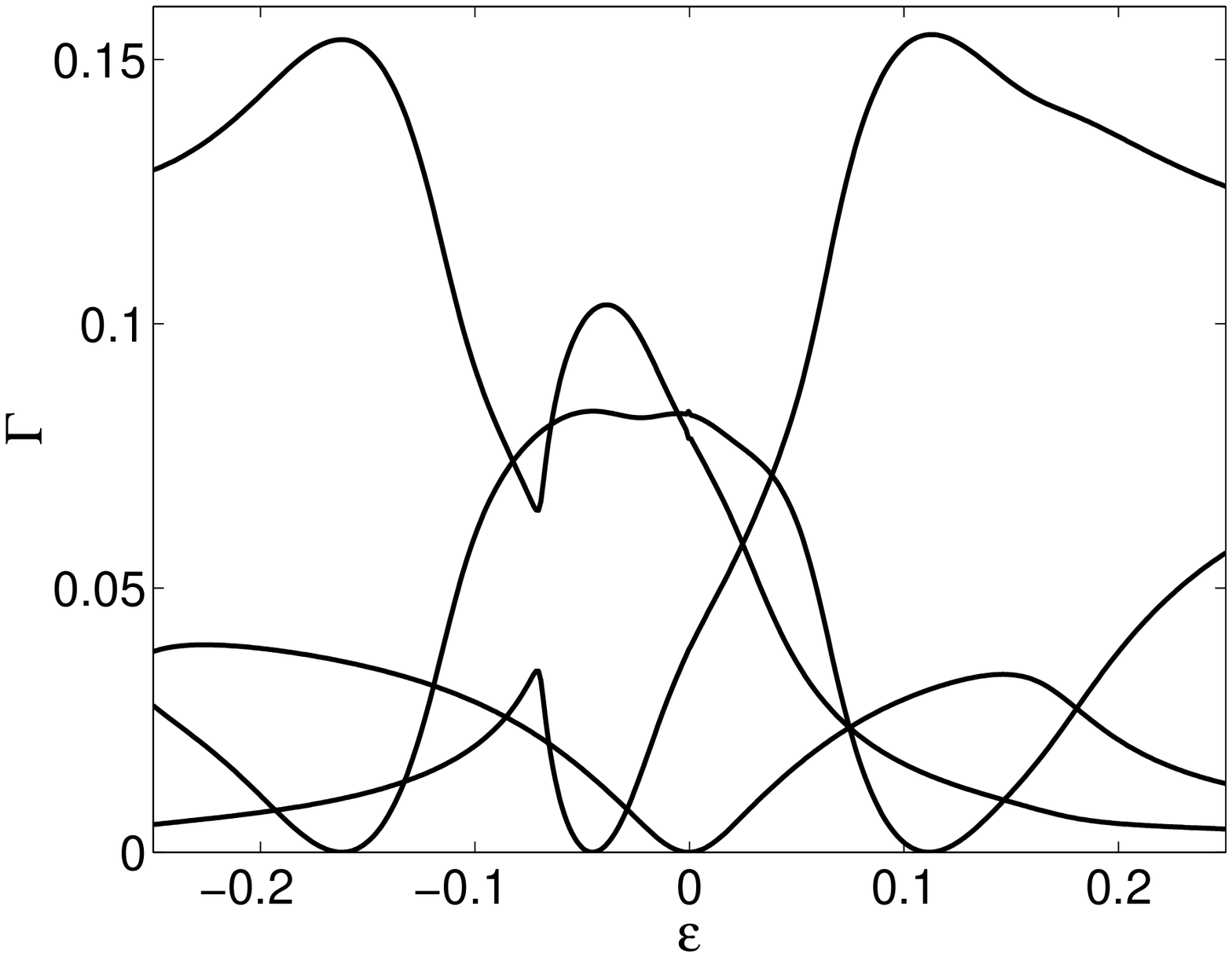}
\caption{(a) The transmission $\ln(-\ln(1-|T|))$ of QD versus
energy of incident electron and energy splitting $\epsilon$ in the
avoiding crossing scenario $v=0.05, U_1=0.2, ~U_2=0.3,
\Gamma_1=\Gamma_2=0.05$. Black regions correspond to those where
the transmission close to zero while the white ones do to those
where the transmission is near unit. Thereby the white regions
follow the  single-electron and two-electron energy levels in
closed QD. Positions of BICs are shown by open circles. (b) The
resonance widths dependent on $\epsilon$.} \label{fig4}
\end{figure}
\cite{volya,PRB} we lift the degeneracy in the QD by transitions
between levels, adding a hopping term between the single-electron
states into the Hamiltonian of the two-level QD, $H_D\rightarrow
H_D-va_{1\sigma}^{+}a_{2\sigma}-va_{2\sigma}^{+}a_{1\sigma}$ which
evolves the picture of energy crossing into a picture with an
avoided crossing. Fig. \ref{fig4} (a) shows the transmission of QD
in which the energy levels repel each other because of the hopping
between QD levels. In order to show clearly the zero and unit
transmission we present the transmission in the double log scale
$\ln(-\ln(1-|T|))$. One can see the avoided level crossings shown
by white lines with $T=1$. BICs shown by open circles are located
at those points where the unit transmission $T=1$ (white lines)
crosses the zero one $G=0$ (black lines) similar to the case of
non interacting electrons \cite{PRB}. Fig. \ref{fig4} (b) shows
that the resonance widths turn to zero at four critical values of
$\epsilon$.

\begin{figure}[ht]
\includegraphics[width=.5\columnwidth]{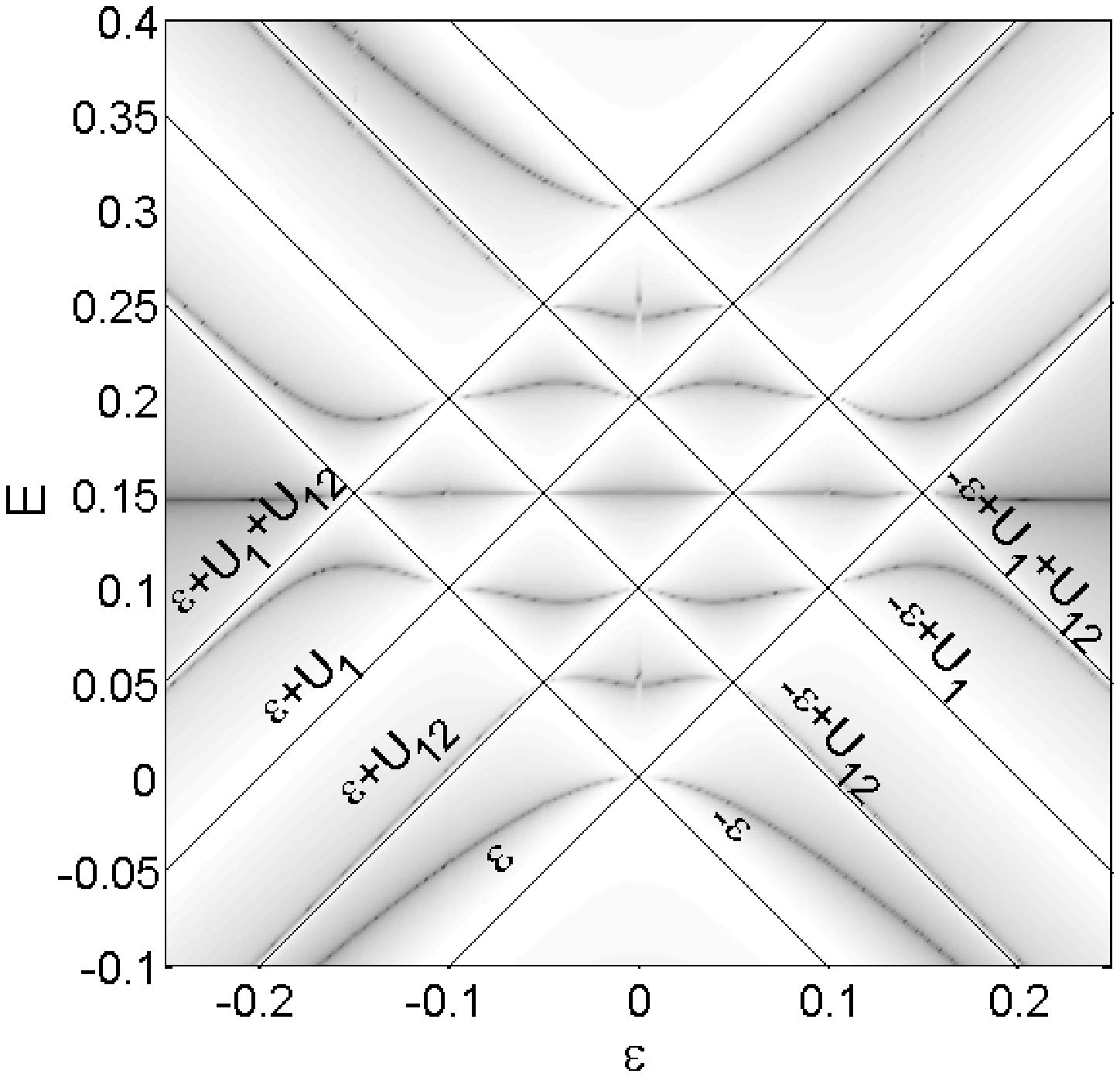}
\includegraphics[width=.44\columnwidth]{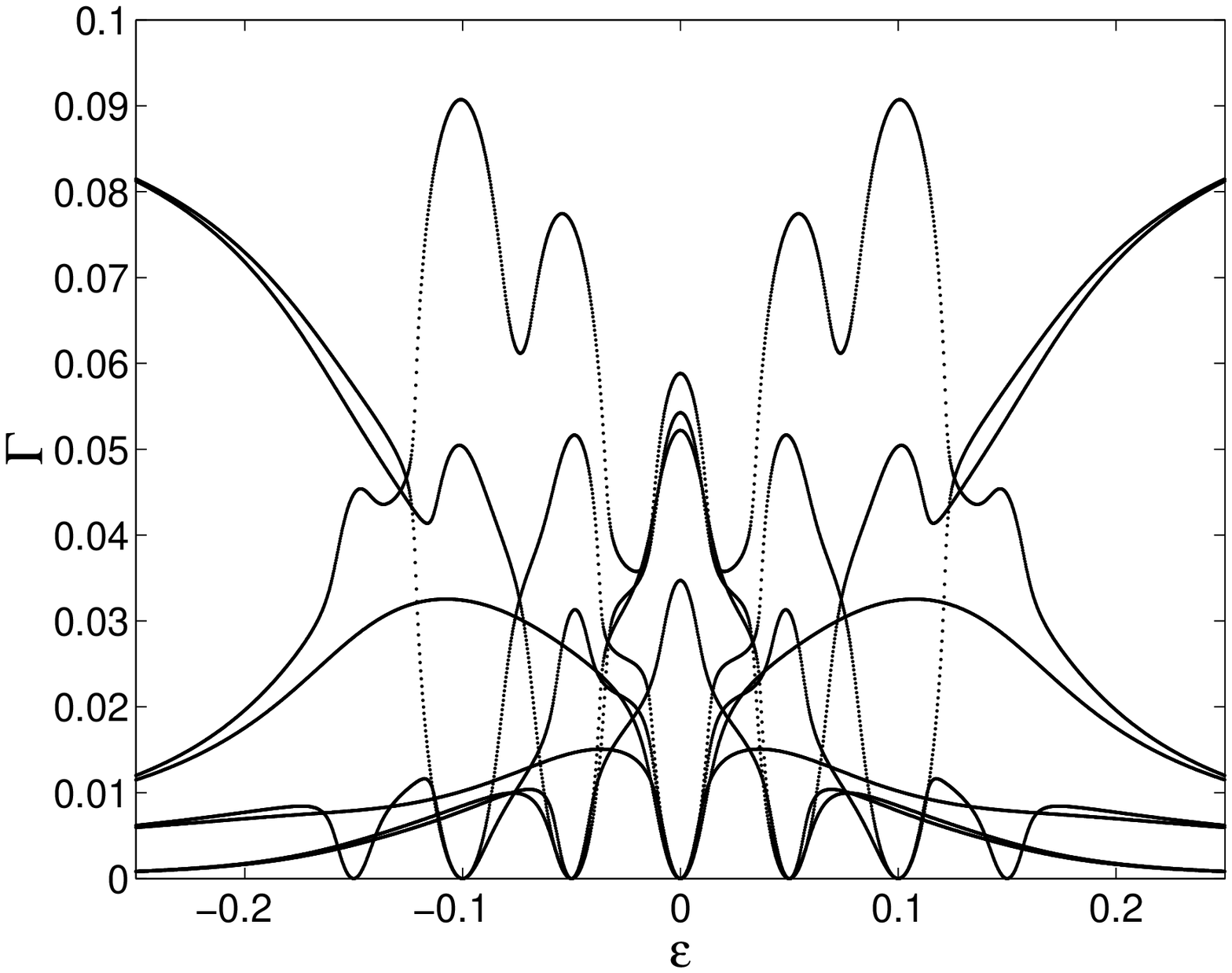}
\caption{The same as in Fig. \ref{fig4} but with inclusion of the
inter-level Coulomb interactions $U_{12}=0.1$.(a) The transmission
$\ln|T|$ of QD versus energy of incident electron and energy
splitting $\epsilon$ in the avoiding crossing scenario $v=0.05,
U_1=U_2=0.2, \Gamma_1=\Gamma_2=0.1$.  (b) The resonance widths
dependent on $\epsilon$.} \label{fig5}
\end{figure}

The Hubbard repulsion presented in the Anderson impurity model
(\ref{QD}) is not the only way to account the Coulomb
interactions. The last also induce the inter-level couplings in
the form
$\sum_{mn}\sum_{\sigma\sigma'}U_{mn}n_{m\sigma}n_{n\sigma'}$.
Therefore, in the two-level approach a new Coulomb constant
$U_{12}$ appears. The equations of motion for the Green functions
in the QD become tedious but still complete to give the following
Green function
\begin{equation}\label{U12}
G_{QD,m\sigma}(E)=\frac{(E+E_m)(E+E_m-U_m-U_{12}\langle
n_{\overline{m}\sigma}\rangle)(E+E_m-U_{12}-U_m\langle
  n_m\rangle)}{(E+E_m-U_m(1-\langle
  n_{m\sigma}\rangle)-U_{12}\langle
  n_{\overline{m}\sigma}\rangle)(E+E_m-U_{12}(1-\langle
  n_{\overline{m}\sigma}\rangle)-U_m\langle
  n_{\overline{m}\sigma}\rangle)(E+E_m-U_m-U_{12}\langle
  n_{\overline{m}\sigma})\rangle)},
  \end{equation}
where $m=1,2, \overline{m}=2,1, ~E_m=\mp \epsilon$. A substitution
of (\ref{U12}) into Eqs. (\ref{Dyson}), (\ref{occup}) and
(\ref{T}) allows to calculate the transmission of the QD presented
in Fig. \ref{fig5} (a).
Each crossing of energy levels shown by solid lines gives rise to
BICs as shown in Fig. \ref{fig5} (b). One can see that for
$\epsilon=0$ there are simultaneously four crossings. As a result
at this points four resonance width turn to zero as shown in Fig.
\ref{fig5} (b). Corresponding at the points $\epsilon=\pm 0.05$ we
obtain three BICs and so on. If to compare all Figures with
transmission probability through the QD one can see that the
Coulomb interactions in QD replicate the transmission zeros which
are between neighboring resonances. If the resonances are crossing
by an effect of gate voltage we observe BICs at each crossing
points \cite{friedrich} as Figs \ref{fig2}, \ref{fig4} (b) and
\ref{fig5} (b) show.

\indent {\bf Acknowledgements} AF thanks Igor Sandalov for helpful
discussions.

\end{document}